\documentclass{iaus}
\usepackage{graphicx}

\title[Wide Binaries and Moving Groups] 
{Halo Wide Binaries and Moving Clusters as Probes of the Dynamical
and Merger History of our Galaxy}

\author[Allen, Poveda, \& Hern\'andez-Alc\'antara]   
{Christine Allen$^1$, Arcadio Poveda$^1$\and A.
Hern\'andez-Alc\'antara$^1$}

\affiliation{$^1$Instituto de Astronom\'ia, Universidad Nacional Aut\'onoma de M\'exico, 
Ciudad Universitaria, 04510 M\'exico D.F.\break
email: chris@astroscu.unam.mx\\[\affilskip]
}

\pubyear{2006}
\volume{240}  

\pagerange{119--126}

\date{?? and in revised form ??}

\jname{Proceedings Title IAU Symposium}

\editors{A.C. Editor, B.D. Editor \& C.E. Editor, eds.}

\begin{document}

\maketitle

\begin{abstract}
Wide or fragile pairs are sensitive probes of the galactic
potential, and they have been used to provide information about
the galactic tidal field, the density of GMC and the masses of
dark matter perturbers present in both the disk and the halo.  Halo wide
binaries and moving clusters, since they are likely to be the
remains of past mergers or of dissolved clusters, can provide
information on the dynamical and merger history of our Galaxy.
Such remnants should continue to show similar motions over times
of the order of their ages.  We have looked for phase space
groupings among the low-metallicity stars of  Schuster \etal\
(2006) and have identified a number of candidate moving clusters.
In several of the moving clusters we found a wide CPM binary
already identified in our catalogue of wide binaries among
high-velocity and metal-poor stars (Allen \etal\ 2000a).
Spectroscopic follow-up studies of these stars would confirm the
physical reality of the groups, as well as allow us to distinguish
whether their progenitors are  dissolved  clusters or accreted
extragalactic systems. 

\end{abstract}

\firstsection 

\section{Introduction}

In recent times evidence has accumulated indicating that the
galactic halo formed at least in part by accretion of material
from smaller extragalactic systems.  This view is in accordance
with the hierarchical theory of structure formation in the
Universe, in which the first objects to form are small galaxies,
which then merge into the larger structures observed today.  A
consequence of this theory is that substructures should still be
present within the old stars of the galactic halo and disk.  Thus,
the merger history of our galaxy should show up in the
distribution of old stars, as coherent tidal streams, tails,
moving clusters, etc. (Majewski 2004, 2005 and references
therein).

But quite apart from the merger history of our galaxy, the stellar
halo we now observe is also the result of the dynamical disruption
of many globular clusters.  Such disrupted clusters should also
leave their signatures in the motions of the halo stars and should
show up as dynamically coherent groupings in phase space.  Poveda
\etal\ (1992) searched for such groupings among the nearby halo
stars of the first Schuster \etal\ catalogues (Schuster and Nissen
1988, 1989a, 1989b, henceforth SN1) and they were able to identify
seven moving groups, characterized by the similarity of their
integrals of motion (the energy  $E$ and the $z$-component of the
angular momentum $h$) as well as by their similar metallicities
[Fe/H].  In particular, the Kapteyn and the Groombridge 1830
moving groups were re-discovered.

On the other hand, there exists a small number of old and
extremely wide binaries ($a>10000$ AU) that are very difficult to
understand dynamically (Allen \etal\ 2000b, Carney \etal\ 1997).
In Table 1 we provide a few examples. The first  six rows list 
parameters of the galactic orbits of these binaries.  The metallicity, 
[Fe/H], is given in row 7.  The last two rows list the observed 
(projected) separations of the binaries and the expected major semiaxes, 
calculated according to the theoretical formula given by Couteau (1960).
The values for the major semiaxes are quite large, and since these 
binaries are old, they should have been already dissociated.  One possibility 
is that they have survived because they spend only a small fraction of their
lives in the galactic disk, and they cross it at high speeds
(Allen \etal\ 2000a).  But perhaps there is another way in which
they can survive to the present day as binaries.  They may, in
fact, have been part of bound moving halo clusters, now dissolving or
already dissolved.


\begin{table}
  \begin{center}
  \caption{Properties of some extremely wide, old binaries}
  \label{}
  \begin{tabular}{lccccc}
  \hline
                                   & W1828       & G15-10      & G40-14    & LDS519      & BD+80$^\circ$245\\
                                   & (Hip 15396) & (Hip 74199) & ($-$)     & (Hip 74235) & (Hip 40068)\\
  \hline
  $E$ (100 km$^2$s$^{-2}$)         & $-$1163.1   & $-$847.3    & $-$1153.7 & $-$524.6    & $-$878.3\\
  $h$ (10 kpc km s$^{-1}$)         & $-$143.1    & $-$41.8     & $-$69.6   & $-$238.1    & $-$118.5\\
  $R_{{\rm min}}$ (kpc)            & 7.6         & 1.4         & 1.5       & 5.0         & 6.0\\
  $R_{{\rm max}}$ (kpc)            & 9.6         & 30.2        & 15.1      & 69.2        & 25.8\\
  $\vert z_{{\rm max}}\vert$ (kpc) & 5.2         & 24.1        & 0.2       & 7.1         & 20.5\\
  Eccentricity                     & 0.12        & 0.91        & 0.81      & 0.87        & 0.62\\
  $[$Fe/H$]$                       & $-$2.05     & $-$2.42     & $-$2.46   & $-$1.28     & $-$2.07 \\
  $s$ (AU)                         & 39590       & 89366       & 23030     & 8846        &  6513 \\
  $\langle a\rangle$ (AU)          & 55410       & 125073      & 32283     & 12380       &  9115\\
  \hline
  \end{tabular}
  \end{center}
  \end{table}

The wide binaries and other substructures observed in the stellar
halo of our galaxy may be the  remains either of past mergers, or
of disrupted clusters.  An interesting question arises:  how can
we distinguish between both possibilities?  Detailed chemical
analyses can provide answers in some cases (King 1997; Ivans
\etal\ 2003;  Venn \etal\ 2004).

In the present paper we look for  moving groups with the same
technique used in Poveda \etal\ (1992),  but we expand the SN1
sample with the more recent catalogue of Schuster \etal\ (2006),
henceforth referred to as SN3.  The total number of stars
contained in the expanded sample is 1451, out of which 483 turn
out to be halo stars according to the $V_{{\rm rot}}$ - [Fe/H]
criterion advocated by Schuster \etal\ to distinguish between
halo, thick disk and disk stars.

The outline of this paper is as follows.  In Section 2 we define
our sample of stars and discuss the technique we used for
identifying moving groups.  Section 3 contains examples of possible
halo moving groups, and compares these groupings with those of other
workers who have used similar techniques.  Section 4 explores the
connexion between our moving groups and some of the old and very
wide binaries.  Sections 5 and 6 study the connexion between
chemically peculiar stars and wide binaries or moving groups.
Section 7 summarizes our main results.



\begin{figure}[!t]
\begin{center}
\includegraphics[width=5.2in,height=7.2in]{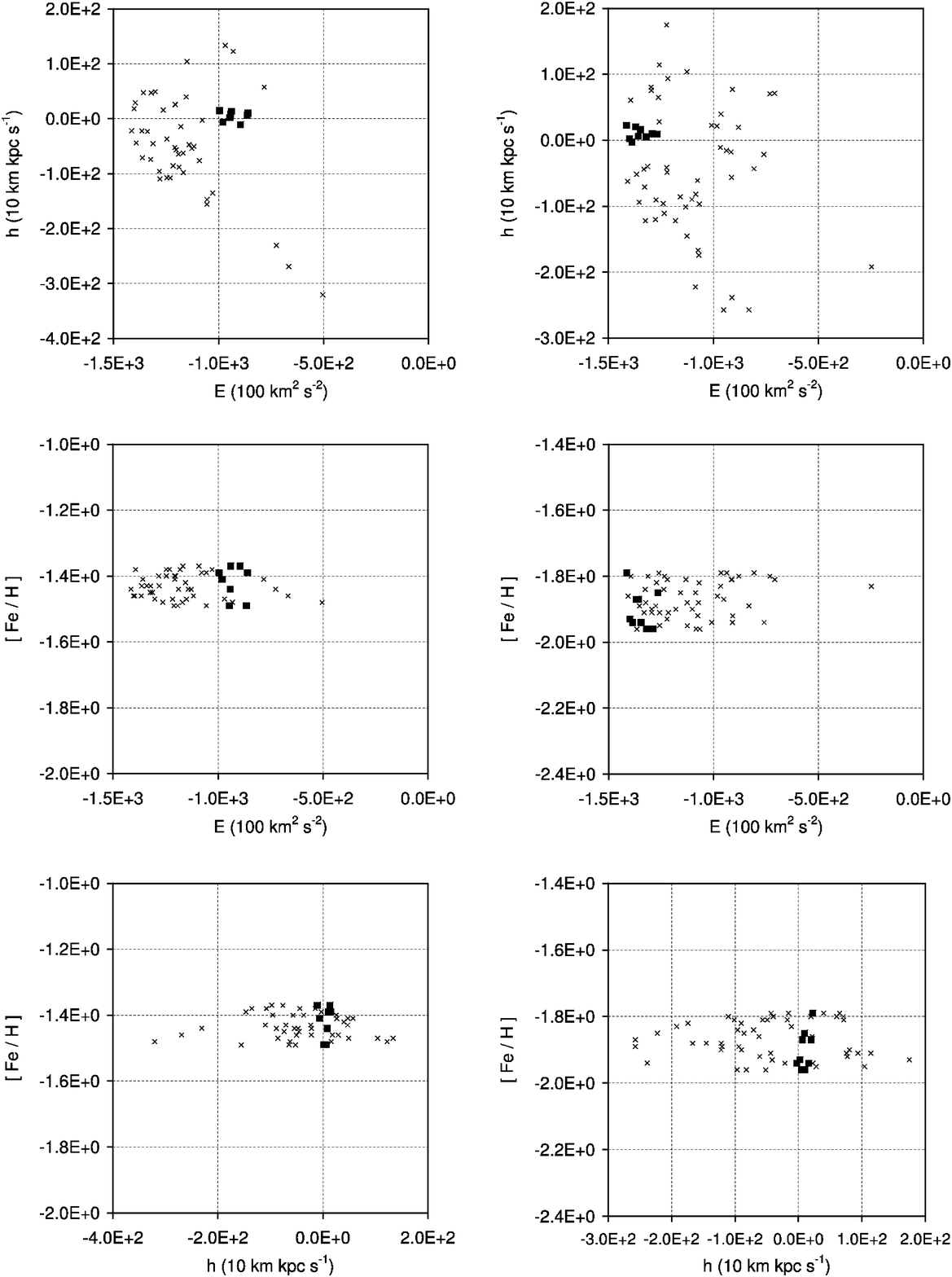}
\end{center}
\caption{Left column. Plots of the two integrals of motion, $E$,
$h$, and of the metallicity, [Fe/H] for the candidate moving
cluster SN8. The stars belonging to this group are plotted as
filled squares.  The Schuster \etal\ (2006) stars with
metallicities similar to those of the group members are plotted as
crosses. Right column. Same plots for the candidate moving cluster
SN11} \label{fig1}
\end{figure}

\section{Identification of moving groups}

To identify moving groups we made use of the existence of two
integrals of motion in an axisymmetric potential, namely the
energy, $E$, and the $z$ component of the angular momentum, $h$.
We computed these quantities for the  halo stars of the combined
sample SN3.  We next looked for groupings in the $E$, $h$ plane,
considering only
stars having similar values of the metallicity [Fe/H].  We will
refer to these groupings with similar $E$, $h$ and [Fe/H] as
``phase space clumps''.   We choose the $E$, $h$ plane as the most
straightforward way of plotting quantities that are strictly
conserved in a time-independent axisymmetric potential. Recent
work (Pichardo \etal\ 2004; Allen \etal\ 2006) has shown that
even in the presence of a bar, those quantities are also conserved
on the average, except for orbits residing entirely within the
bar, which is certainly not the case for relatively nearby stars
such as the SN3 sample. Therefore, $E$ and $h$ will keep memory for
long times of the values they had when they formed a group.  We
also require similar values of the metallicity, since the stars
within a former group are likely to have but a small spread in
their chemical compositions. A further requirement to identify
group members is the similarity of their galactic orbital
parameters, especially the maximum $\vert z\vert$-value.


\begin{table}
  \begin{center}
  \caption{Group 8 - members}
  \label{2}
  \begin{tabular}{lccccccc}\hline
    Leading   &     $E$            & $h$               & [Fe/H] & $\vert z_{{\rm max}}\vert$
 & $R{{\rm max}}$ & $R{{\rm min}}$ & $e$ \\
           star name       & 100~km$^2s^{-2}$ & 10~km kpc s$^{-1}$ &          &  kpc &  kpc & kpc\\
 \hline
 G192-043 & $-$895.8 & $-$10.8 & $-$1.37 & 17.4 & 27.8 & 0.16 & 0.99 \\
 G088-023 & $-$995.4 &    14.7 & $-$1.39 & 17.8 & 21.8 & 0.25 & 0.98 \\
 G202-065 & $-$879.9 &  $-$6.4 & $-$1.41 & 13.7 & 22.8 & 0.09 & 0.99 \\
 G018-024 & $-$946.4 &     2.0 & $-$1.49 & 22.9 & 23.1 &  1.4 & 0.89 \\
 G179-054 & $-$942.7 &     8.1 & $-$1.44 & 20.8 & 24.9 &  0.2 & 0.99\\
 G215-047 & $-$939.6 &    13.5 & $-$1.37 & 18.2 & 24.9 &  0.2 & 0.98\\
 G072-006 & $-$864.4 &     6.4 & $-$1.49 & 23.7 & 29.9 &  0.1 & 0.99 \\
 G126-063 & $-$860.0 &    10.1 & $-$1.39 & 23.1 & 30.1 &  0.2 & 0.99 \\
 \hline
  \end{tabular}
 \end{center}
\end{table}


\begin{table}
  \begin{center}
  \caption{Group 11 - members}
  \label{3}
  \begin{tabular}{lccccccc}\hline
 Leading      &     $E$            & $h$               & [Fe/H] & $\vert z_{{\rm max}}\vert$
 & $R_{{\rm max}}$ & $R_{{\rm min}}$ & $e$ \\
star name       & 100~km$^2s^{-2}$ & 10~km kpc s$^{-1}$ &          &  kpc &  kpc & kpc\\
\hline
 LP720-028  & $-$1399.9  &    2.3  & $-$1.93  & 6.7  &  9.1 & 0.04 & 0.99 \\
 HD116064   & $-$1320.2  &    4.9  & $-$1.96  & 8.3  & 10.7 & 0.08 & 0.99 \\
 LP770-071  & $-$1359.0  &    6.5  & $-$1.87  & 7.4  &  9.9 & 0.11 & 0.98 \\
 LHS2969    & $-$1267.3  &    9.6  & $-$1.85  & 8.2  & 11.9 & 0.16 & 0.97 \\
 G029-071   & $-$1290.4  &    9.8  & $-$1.96  & 7.9  & 11.4 & 0.16 & 0.97\\
 LP673-106  & $-$1347.1  &   15.9  & $-$1.94  & 7.4  &  8.9 & 0.83 & 0.83\\
 G088-027   & $-$1370.0  &   20.3  & $-$1.87  & 6.3  &  9.6 & 0.38 & 0.92 \\
 HD160617   & $-$1387.8  & $-$2.9  & $-$1.94  & 7.4  &  9.3 & 0.05 & 0.99 \\
 G130-007   & $-$1414.0  &   22.4  & $-$1.79  & 5.4  &  8.8 & 0.41 & 0.91\\
  \hline
  \end{tabular}
 \end{center}
 \vspace{0.3cm}
\end{table}


\begin{figure}[!t]
\begin{center}
\includegraphics[height=7.2in,width=5.2in]{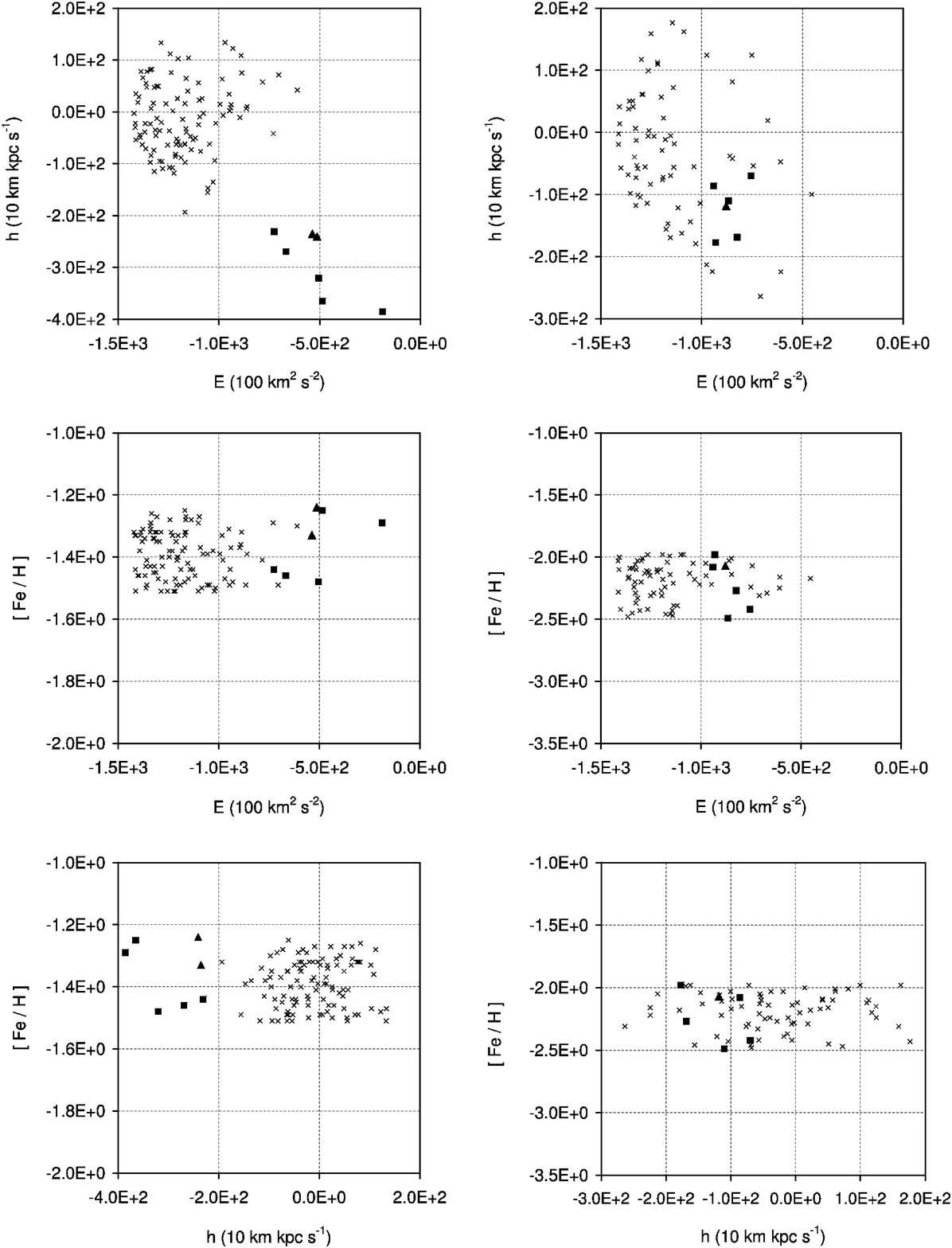}
\end{center}
\caption{Left column. Plots of the two integrals of motion, $E$,
$h$, and of the metallicity, [Fe/H] for the candidate moving
cluster associated with the wide binary HD~134439/40.  The stars
belonging to this group are plotted as filled squares. The
Schuster \etal\ (2006) stars with metallicities similar to those
of the group members are plotted  as crosses. Right column.  Same
plots for the candidate moving cluster associated with the wide
binary BD+80$^\circ$245}\label{fig2}
\end{figure}

As a criterion to define ``similar values'' we take the case of
the well-studied common proper motion binary LDS 519
($=$ HIP 74234/74235 $=$ HD134439/134440).  For this pair, we 
can compute the difference
of the energies and $z$-components of the angular momenta of the 
galactic orbit of each component, as well as the difference in metallicity. Since
there is little doubt that both components of  this wide pair are
physically connected, we can take these differences as
representative of the differences we would expect to find among
members of a common group.

\section{Halo moving groups among the SN3 stars}

We used the halo stars extracted from the Schuster \etal\
catalogues selecting them according to the population criterion
$X$ advocated by these authors, which is a linear combination of
the rotational velocity, $V_{{\rm rot}}$, and the metallicity,
[Fe/H]. A positive value of $X$ identifies a halo star.

In this way, we ended up with 478 stars belonging to the halo
population. For these stars we computed galactic orbits using the
Allen \& Santill\'an (1993) axisymmetric model potential.  We
looked for groupings in this halo star sample in the manner
described above, and we were able to identify about a dozen
candidate moving clusters, which satisfy the criteria of having
similar values of $E$, $h$, [Fe/H], and similar orbital
parameters. A full discussion of these halo moving groups will be
given elsewhere. Here, we only show a few  examples. Tables 2 and
3 list the members of two candidate groups.  Successive rows list the 
galactic orbital parameters.  Figure 1 displays plots of their member 
stars in three planes containing combinations of
$E$, $h$, and  [Fe/H].  Note that the groupings are much tighter
(especially in the planes containing [Fe/H]) than similar groupings
proposed in the literature (Eggen 1996, Fiorentin \etal\ 2004,
Chiba \& Beers 2000, Helmi \& White 1999).

\section{Wide binaries among moving groups}

Interestingly, we find wide binaries in 6 of our candidate moving
clusters.  These binaries were found by searching among the  wide
binaries of the catalog of Allen \etal\ (2000a), which was
constructed using earlier versions of the Schuster \etal\
catalogues. The binaries were found independently of our search for
moving clusters.  The fact that some moving groups contain wide
binaries supports our conjecture that some of the old wide
binaries may be the survivors of former moving groups, which have
already dissolved or are in the process of doing so.  These
binaries are listed in Table 4.  The rows and units in this table are 
the same as in Table 1.

Among the old wide binaries listed in Table 1 there are two
well-studied objects that have very peculiar chemical abundances,
namely LDS 519 and BD+80$^\circ$245 (Chen \& Zhao 2005, King 1997,
Zhang \& Zhao 2005, Ivans 2003, Zapatero-Osorio \& Martin 2004).
Specifically, they show very low ratios of [$\alpha$/Fe] for their
metallicities [Fe/H]. Such chemical peculiarities point to an
enrichment history different from that of the bulk of our galactic
halo.  In fact, it has  been suggested that they are reliable
indicators of accreted structures. LDS 519 is contained in the SN3
catalog, and its galactic orbit was obtained with the parameters
given in that source. For BD+80$^\circ$245, not contained in the
SN3 catalog, we were able to obtain an orbit using data on its
distance and kinematics from Hipparcos and Carney \etal\ (1997).

\section{Moving groups around chemically peculiar stars}

We may ask, conversely, whether there is any evidence of stars
with values of $E$, $h$ and [Fe/H] similar to those of these two
chemically peculiar stars.  We emphasize that they are both 
members of wide binaries. Our findings are displayed in Figure 2. 
We can see that there are indeed other stars with values of
$E$, $h$, [Fe/H] that are similar to those of LDS 519 and
BD+80$^\circ$245, although the groupings are not as tight as the
ones we showed before.  No detailed chemical abundances (ie,
[$\alpha$/Fe] ratios) are available for these  candidate moving
cluster members, but it would be very interesting if they should
also turn out to be peculiar. This would confirm their membership
to the halo moving clusters here proposed, and constitute evidence
for their origin by accretion.

\section{Binaries or moving cluster members with peculiar chemical
composition}

Another question we can ask is whether there are other binaries or
moving groups with peculiar values of [$\alpha$/Fe].  To find an
answer, we made use of the detailed study of Venn \etal\ (2004).
We looked  in their Table 2 for information on [$\alpha$/Fe]  for
all the wide binaries of our catalog (Allen et al. 2000a), as well
as for the candidate members of moving groups we identified among
the SN3 stars.  The result of this search is shown in Figure 3,
adapted from Venn et al. (2004).  We show as crosses the stars and
as hollow squares the dwarf galaxies studied by Venn \etal\.  The
wide binaries of Allen \etal\ appear plotted as thick dots, and
the members of candidate moving groups as triangles.  The figure
shows clearly that four of the stars identified as possible moving
group members, namely HD 105004, HD 134439/40 (LDS519) and
BD+80$^\circ$245 clearly have low [$\alpha$/Fe] for their [Fe/H].
Recall that BD+80$^\circ$245 is itself a wide binary, although only the 
primary is plotted in Figure 3. Such low [$\alpha$/Fe] stars are chemically 
similar to the dwarf galaxies, and are therefore probably accreted 
structures.  Figure 3 also shows that at least five of our old wide 
binaries also have such chemical peculiarities, and hence may also be 
the remains of accreted dwarf galaxies.


\begin{figure}[!t]
\begin{center}
\includegraphics[height=2.5in,width=5.3in]{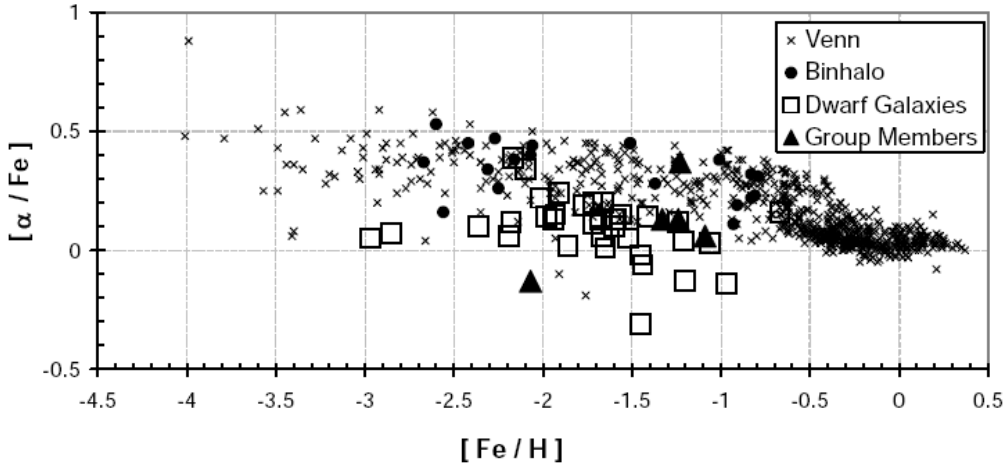}
\end{center}
\caption{Plot of the abundance of alpha elements vs metallicity
(adapted from Venn \etal\ 2004). The crosses  denote stars, the empty
squares, dwarf galaxies, as plotted by Venn \etal\. We have added as dots
the wide binaries of Allen \etal\ (2000a) for which alpha-element
abundances were available in Venn \etal\.  Triangles denote binaries that
are members of  moving clusters.}\label{fig3}
\end{figure}


\begin{table}
  \begin{center}
  \caption{Binaries in moving clusters}
  \label{4}
  \begin{tabular}{lccccccc}\hline
    &  HD~105004  & HD~23439  & G18-24 & G40-14 & G15-10  & HD~134439/40 & BD+80$^\circ$245 \\
 \hline
 $E$                        & $-$1382.5 & $-$1304.3  & $-$946.4  & $-$1153.7  & $-$847.3 &$-$524.6  &$-$878.3 \\
 $h$                        & $-$19.8   &      99.4  &      2.0  &   $-$69.6  &  $-$41.8 &$-$238.1  &$-$118.5 \\
 $R_{{\rm min}}$            &     0.3   &       3.1  &      1.4  &       1.5  &      1.4 &     5.0  &    6.0 \\
 $R_{{\rm max}}$            &     9.3   &       9.6  &     23.1  &      15.1  &     30.2 &    69.2  &   25.8 \\
 $\vert z_{{\rm max}}\vert$ &     6.1   &       1.5  &     23.0  &       0.2  &     24.1 &     7.1  &   20.5 \\
 $e$                        &     0.93  &       0.52 &      0.89 &       0.81 &     0.91 &     0.87 &    0.62\\
 $[$Fe/H$]$                 &  $-$1.09  &    $-$1.23 &      1.49 &    $-$2.46 &  $-$2.42 &  $-$1.28 & $-$2.07\\
 $s$                        &  1620     &   196      &   11782   &   23030    &   89366  &   8846   & 6513\\
 $\langle a\rangle$         &  2267     &   274      &   16489   &   32233    &  125073  &  12380   & 9115\\
       \hline
  \end{tabular}
 \end{center}
\end{table}

\section{Summary and conclusions}

Old wide binaries may have survived either because they spend
little time in the galactic disk, crossing it at high speeds, or
else because they were once members of bound clusters or still 
belong to moving groups. A search for ``phase space clumps'' 
among the stars of Schuster
\etal\ (2006) yielded about a dozen groups of similar
metallicities and galactic orbits.  Previously known wide binaries
turned out to be present in six of the identified moving groups.
We may turn to detailed chemical composition analyses to
distinguish whether a grouping had a galactic  origin or else if
it is an accreted structure. Chemical peculiarities, like low
[O/Fe] or especially, low [$\alpha$/Fe] are characteristic of
enrichment by type Ia supernovae, and thus of a chemical history
different from that of our galaxy. Along with high eccentricities
of the galactic orbits, or retrograde motions, these chemical
peculiarities are thought to be signatures of accreted structures
(dwarf galaxies).  At least 5 of our old wide binaries have such
peculiarities, and 3 are members of moving groups, which thus may
be the remains of dwarf galaxies accreted onto the galactic halo.

\begin{acknowledgments} Our thanks are due to W. J. Schuster for fruitful 
discussions.  C.A. is grateful to the IAU for a travel grant.

\end{acknowledgments}

\end{document}